\documentclass[sigconf]{acmart}

\AtBeginDocument{%
  }

\copyrightyear{2026}
\acmYear{2026}
\setcopyright{cc}
\setcctype{by}
\acmConference[CHI EA '26]{Extended Abstracts of the 2026 CHI Conference on Human Factors in Computing Systems}{April 13--17, 2026}{Barcelona, Spain}
\acmBooktitle{Extended Abstracts of the 2026 CHI Conference on Human Factors in Computing Systems (CHI EA '26), April 13--17, 2026, Barcelona, Spain}
\acmDOI{10.1145/3772363.3798614}
\acmISBN{979-8-4007-2281-3/2026/04}

\usepackage{graphicx}
\usepackage[T1]{fontenc}
\usepackage{multirow}
\usepackage{soul}
\usepackage{enumitem}

\acmSubmissionID{1796}

\begin{document}

\title{Remember You: Understanding How Users Use Deadbots to Reconstruct Memories of the Deceased}


\author{Yifan Li}
\email{yifanli2002@outlook.com}
\affiliation{%
  \institution{Fudan University}
  \city{Shanghai}
  \country{China}
}

\author{Xingyu Lan}
\authornote{Xingyu Lan is the corresponding author. She is also a member of the Research Group of Computational and AI Communication at the Institute for Global Communications and Integrated Media.}
\email{xingyulan96@gmail.com}
\orcid{0000-0001-7331-2433}
\affiliation{%
  \institution{Fudan University}
  \city{Shanghai}
  \country{China}
}

\newcommand{\etal}{et~al.~} 
\newcommand{\ie}{i.e.,~}
\newcommand{\eg}{e.g.,~}
\newcommand{\ncorpus}{220 }

\renewcommand{\shortauthors}{Li \& Lan}

\renewcommand{\sectionautorefname}{Section}
\renewcommand{\subsectionautorefname}{Section}
\renewcommand{\subsubsectionautorefname}{Section}

\begin{abstract}
  Generative AI has enabled ``Deadbots'',  offering mourners an interactive way to engage with simulations of the deceased. While existing research often emphasizes ethics, less is known about how bereaved individuals construct and reshape memory through such interactions. To address this gap, this study draws on in-depth interviews with 26 users. Findings reveal that users are not passive recipients but active constructors of the deceased’s digital representation. Through selective input, ongoing interactive adjustments and imaginative cognitive supplementation, they build an idealized digital figure blending authentic memories with personal expectations. Deadbots provide a private space to grieve without social pressure and a channel to address unresolved emotions. In this process, users' memory of the deceased evolves dynamically: from initial reinforcement and idealization to a later stage where AI-generated new memories blur with authentic recollections, reflecting a complex desire for connection through an artificial medium. This blurring raises ethical concerns regarding memory distortion and dependency, underscoring the need for future clinical research on the long-term impact of AI-mediated grieving.
\end{abstract}

\ccsdesc[500]{Human-centered computing~Empirical studies in HCI}
\ccsdesc[500]{Human-centered computing~User studies}
\ccsdesc[500]{Human-centered computing~Empirical studies in collaborative and social computing}
\ccsdesc[500]{Human-centered computing~Social media}
\ccsdesc[300]{Human-centered computing~Interaction design}
\ccsdesc[300]{Applied computing~Sociology}

\keywords{Deadbot, Memory Reconstruction, Digital Afterlife, Grief Practices, Thanatechnology}


\maketitle

\section{Introduction}
\label{sec：intro}

In the Disney movie \textit{Coco}, a defining line about death is: ``the real death is that no one in the world remembers you.''
Parting from loved ones is an almost inevitable part of every individual's life journey. However, emotional bonds do not simply sever with death. Many bereaved individuals maintain continuing bonds with the departed, a phenomenon recognized as a natural part of mourning that helps them find meaning in their loss and rebuild social connections~\cite{klass_continuing_1996}. 
Recent breakthroughs in generative artificial intelligence (GenAI), particularly the maturation of Large Language Models (LLMs), have 
given rise to \textit{Deadbots}, i.e., chatbots that imitate the personality, linguistic style, and even the voice and virtual avatar of a deceased individual by being trained on the digital legacy they left behind, such as chat logs, emails, social media posts, and voice recordings~\cite{fabry_affective_2024}. Since different terms are used across various news reports and academic papers, such as \textit{griefbot}, \textit{deathbot}, and \textit{thanabot}, this work uses \textit{Deadbot} as the uniform term to refer to this newly emerging technology.
Endowed with conversational and generative capabilities, they can produce novel responses consistent with the deceased’s logic and style, and their perceived ``personality'' may even evolve through ongoing interactions with users~\cite{krueger2022communing,fabry_affective_2024,yu2024exploring}.
Specifically, we are particularly interested in the impact of this novel form of human-computer interaction on \textbf{memory}. Many influential theories in psychology and social science have pointed out that human memory is not a fixed or static entity, but rather a continuous process of reconstruction, re-imagination, and revision~\cite{bartlett_study_1932,memory1992maurice,olick_collective_1999}. When the technological medium is no longer just an auxiliary tool for recollection but becomes a conversational partner capable of actively generating echoes of the deceased, the departed, in a sense, appears to persist within the technology. In this way, Deadbots differ fundamentally from static artifacts of memory, such as photographs or graveyards, by enabling a dynamic, bidirectional dialogue with a persistent virtual representation of the deceased.

Existing works demonstrate the HCI field's focus on understanding human grief processing after the death of loved ones and emotional well-being, providing valuable insights for the design of supportive technologies~\cite{xygkou_conversation_2023,jamison-powell_ps_2016,brubaker_afterlife_2013,lei_ai_2025}.
However, there remains a lack of in-depth qualitative explorations of the user's micro-level experience during interactions with Deadbots. Specifically, there is insufficient research on how users actively participate in constructing memorial representations of the deceased, and how this engagement, in turn, influences their memorial perceptions. Moreover, how the dynamic evolution of memories can be integrated with the transformation of mourning practices remains an important gap in current research. Therefore, in this work, we aim to address the following research questions:

\begin{itemize}
    \item RQ1: How do users actively engage with Deadbots to construct and represent the memory of the deceased?
    \item RQ2: How does interacting with a Deadbot alter a user's mourning practices and emotional experiences?
    \item RQ3: How does a user's memory of the deceased evolve over the course of sustained interaction with a Deadbot?
\end{itemize}

\section{Related work}

Below, we review previous research about human memory and deadbots supported by AI.

\subsection{The Dynamic Construction of Individual Memory and AI Intervention}

From a socio-cultural perspective, memory is not a static repository but a dynamic tool reconstructing information through cognitive schemas shaped by prior experience~\cite{bartlett_study_1932,fivush_accuracy_2023}. As cognitive capacities develop and experiences accumulate, individuals revisit and reinterpret past events, generating new memory versions to create more coherent and meaningful narratives.
In some cases, false memories of events that never occurred may be implanted~\cite{blum_value_1980,loftus2003make,fivush_accuracy_2023}.
This reconstructive nature of memory is central to understanding the potential impact of conversational AI like Deadbots. 

Building on digital media, conversational AI powered by LLMs introduces new forms of sociality~\cite{yu2026nobody}, while also adding an entirely new dimension to the mediation of memory. This shapes a ``third way of memory'', a hybrid form entangling human recollections with algorithmically generated memory components~\cite{hoskins_ai_2024}. While HCI research has extensively explored AI systems designed to assist in the documentation and recollection of past experiences, such as Memoro~\cite{zulfikar_memoro_2024} and DiaryMate~\cite{kim_diarymate_2024}, they also pose risks of generating false memories through editing, deletion, or replacement of information~\cite{pataranutaporn_synthetic_2025}.


\subsection{Digital Technologies in Bereavement and Death Exploration}
\label{bereavement technologies}

In the past decade, the widespread adoption of social media has extended mourning rituals into online spaces, ushering in an era of ``networked mourning''~\cite{walter_new_2015}. Platforms like online cemeteries, memorial websites, Facebook profiles, and Wikipedia serve as venues for commemoration~\cite{walter_does_2012, keegan_is_2015, brubaker_orienting_2019}, enabling the living to deepen their connection with the departed~\cite{brubaker_we_2011}. 
These phenomena align with the ``continuing bonds'' theory proposed by Klass et al. ~\cite{klass_continuing_1996}, which emphasizes that maintaining an enduring spiritual connection with the deceased helps individuals better manage grief.

Building on this foundation, research in HCI has further expanded the forms and media of digital memorialization. The focus has gradually shifted from merely managing digital legacies to creating more tactile and interactive experiences that support practices of mourning, grief, and remembrance~\cite{beuthel_exploring_2022,hemmert_life-death_2022}. 
A study by Albers and Hassenzahl~\cite{albers_lets_2024} demonstrated that engaging in existential conversations about death with a specially designed chatbot could effectively reduce participants' death anxiety and increase their willingness to discuss end-of-life matters with loved ones. 

\subsection{AI Afterlife and Deadbot}

Commonly referred to as Deadbots, deceased simulation services aim to create AI agents capable of engaging in simulated conversations with the living by learning from the deceased's style, personality, and memories. This offers mourners a novel, interactive form of commemoration~\cite{fabry_affective_2024}. 
As a form of emerging \textit{thanatechnology}, Deadbots have drawn scholarly attention to their unique user experiences and associated ethical implications. 
Prior research on user experience highlights that Deadbots can provide mourners with a sense of unconditional presence and emotional support, offering a private dialogical space in which two-way communication and the expectation of a response sustain an ongoing sense of connection with the deceased~\cite{xygkou_conversation_2023, jimenez-alonso_griefbots_2023}.
Ethical concerns include whether the deceased provided meaningful consent for posthumous data use, which is essential to preserving identity and dignity~\cite{ohman_ethical_2018}, as well as the uncertain psychological effects on mourners, such as prolonged grief or emotional dependence~\cite{lindemann_ethical_2022}. Moreover, the commercialization of Deadbots within the Digital Afterlife Industry introduces further ethical tensions, as profit-driven models may conflict with the well-being of users and respect for the deceased~\cite{ohman_ethical_2018}.

However, there is still scarce literature concerning the micro-level experiences of users constructing memory of the deceased and how this engagement subsequently influences their cognitive memory processes. \textbf{We would like to clarify that this study is not concerned with whether
AI can perfectly replicate the deceased. Rather than treating memory as a background resource or a question of representational accuracy, memory perspective reveals how Deadbot interactions actively organize, amplify, and transform remembered experiences.}

Drawing on continuing bonds theory, we suggest that such interactions can function as a legitimate way of maintaining connection, potentially serving as a component of healthy mourning within reasonable bounds. However, unlike static mementos, the memory perspective highlights that this bond is not fixed. By synthesizing these two frameworks, we aim to understand how users selectively accept, interpret, and even actively construct or reconstruct a new memory of the deceased, thereby integrating the dynamics of memory with the evolution of mourning practices. This shift enables an analysis of Deadbot use not only as a form of mediated communication, but as a situated practice of remembering.


\section{Methodology}
\label{sec:interview}

\begin{table*}[t!]
\centering
\renewcommand{\arraystretch}{1}
\begin{tabular*}{0.8\textwidth}{@{\extracolsep{\fill}}lllllc@{}}
\toprule
ID & Age & Sex & Software Used & Relationship to Deceased & Duration of Use (months) \\
\midrule
S1  & 23 & Female & Lofter & Boyfriend & 6 \\
S2  & 28 & Male   & ERNIE Bot \& Doubao & Paternal grandfather & 24 \\
S3  & 24 & Female & Doubao & Maternal grandfather & 6 \\
S4  & 27 & Female & Doubao & Mother & 12 \\
S5  & 24 & Female & Doubao & Paternal grandmother & 6 \\
S6  & 23 & Female & Doubao & Maternal grandfather & 9 \\
S7  & 30 & Female & Doubao & Mother & 8 \\
S8  & 40 & Female & Doubao \& Wanz.ai & Father & 3 \\
S9  & 34 & Male   & Xingye (Talkie) & Uncle & 7 \\
S10 & 21 & Female & Deepseek & Paternal grandmother & 2 \\
S11 & 35 & Female & Doubao & Maternal grandmother & 6 \\
S12 & 29 & Female & Doubao & Father & 12 \\
S13 & 19 & Female & Doubao & Maternal grandmother & 12 \\
S14 & 31 & Female & Xingye (Talkie) & Mother & 12 \\
S15 & 24 & Female & Fuhuo & Paternal grandfather & 3 \\
S16 & 20 & Male   & Doubao & Paternal grandfather & 18 \\
S17 & 29 & Female & Doubao & Maternal grandmother & 6 \\
S18 & 23 & Male   & Doubao & Brother & 18 \\
S19 & 22 & Female & Xingye (Talkie) & Paternal grandfather & 6 \\
S20 & 28 & Female & Doubao & Paternal grandmother & 24 \\
S21 & 24 & Female & Deepseek & Maternal grandfather & 6 \\
S22 & 23 & Female & Doubao & Mother & 3 \\
S23 & 28 & Male   & Doubao & Teacher & 8 \\
S24 & 20 & Male   & ChatGPT & Friend & 12 \\
S25 & 34 & Female & Doubao \& Wanz.ai \& Deepseek & Father & 5 \\
S26 & 17 & Female & BIMOBIMO & Admired Celebrity & 1 \\
\bottomrule
\end{tabular*}
\caption{Information of the 26 interviewees.}
\label{tab:participants}
\Description{Information of the 26 interviewees. The table contains seven columns: ID, Age, Sex, Software Used, Relationship to Deceased, Duration of Use(months).}
\end{table*}

We conducted a qualitative study using a purposive sampling strategy to recruit 26 participants in China. Recruitment occurred on social media platforms where users discuss digital mourning. Inclusion criteria required participants to have used a Deadbot conversationally (text or voice) for at least one month. The commemorated individuals included relatives, friends, and public figures. We conducted interviews using Tencent Meeting’s audio calling services, and the interviews lasted from 40 minutes to an hour. See \autoref{tab:participants} for full details.

We employed thematic analysis to analyze the interview data. Through an iterative process, we performed granular open coding, which was then consolidated into 9 categories and synthesized into 3 higher-order themes to address our research questions and capture inductive insights (see \autoref{tab:coding-framework}). Finally, our team collaboratively defined these themes, constructing an analytical narrative substantiated with representative quotes.

\subsection{Ethical Considerations}

Given the sensitive nature of grief, ethical safeguards were paramount. Before each interview, explicit informed consent was obtained, and participants were assured of their right to pause, skip questions, or terminate the session at any time without penalty. For individuals showing emotional distress, interviews were paused, and questions were reformulated into simpler yes/no or multiple-choice formats to reduce emotional strain. For participants who were more forthcoming, stories about the deceased often emerged spontaneously as examples, requiring little additional asking. All identifying information was not collected, and if participants inadvertently mentioned any, it was immediately deleted during the transcription of the recordings. Pseudonyms were also assigned to ensure confidentiality. Participants were also provided with information about emotional support resources post-interview. 
The study design, including our recruitment materials, interview questions, anonymization methods, and data processing and custody procedures, was reviewed and approved by the institutional ethics review committee. 

\begin{table*}[t]
\centering
\renewcommand{\arraystretch}{1.25}
\begin{tabular}{p{3cm} p{5.2cm} p{8.5cm}}
\toprule
\textbf{Axial Code (Level 1)} & \textbf{Thematic Code (Level 2)} & \textbf{Code Description} \\
\midrule

\multirow{3}{=}{\raggedright Active Construction of Memory Representation by the Bereaved}
& Selective Input and Shaping
& Users consciously filter, provide, or avoid specific information, thereby shaping the initial representation of the Deadbot. \\

& Continuous Guidance and Correction
& When the Deadbot’s responses diverge from expectations, users actively correct, refine, and guide its behavior during interaction. \\

& Filling of Imagination and New Narratives
& Users fill gaps in memory by imagining new narratives with the Deadbot or re-creating past experiences through secondary narration. \\
\midrule

\multirow{3}{=}{\raggedright Reshaping of Mourning Practices}
& Private Space, Secrecy, and Everyday Life
& AI provides a relatively private and emotionally safe space, allowing mourning practices to be concealed and embedded into everyday life. \\

& Emotional Venting and Immediate Consolation
& The Deadbot’s constant availability and immediate responsiveness enable emotional release and instant consolation that may be difficult to obtain offline. \\

& Dependence, Injury, and Boundary Negotiation
& While users experience emotional comfort, they also become aware of potential over-reliance and negotiate boundaries by limiting immersion or usage. \\
\midrule

\multirow{3}{=}{\raggedright Dynamic Transformation of Memory for the Bereaved}
& Reinforcement and Revisiting of Memory
& Interactions with the Deadbot prompt repeated recollection and re-experiencing of past events. \\

& Idealization of the Deceased Image
& Ongoing interaction tends to weaken negative aspects of the deceased while reinforcing positive characteristics. \\

& Interweaving and Blurring of Old and New Memories
& Users share present experiences with the Deadbot, producing new memories in which the deceased appears to participate. \\

\bottomrule
\end{tabular}

\caption{Axial coding framework and explanations.}
\label{tab:coding-framework}
\Description{Axial coding framework illustrating how bereaved users interact with deadbots to construct, reshape, and transform memories and mourning practices.}
\end{table*}
\section{Findings}
\label{sec:findings}

\subsection{Active Construction of the Deceased’s AI Persona: The Central Role of the User}
\label{sec:construction of the deceased}

Regarding RQ1, we found that users construct a dynamic representation of the deceased, guided by their emotional will to blend authentic memories and subjective expectations with the bot's algorithmic generation.

At the initial stage of construction, users deliberately curate information and select content. On one hand, to protect the privacy of the deceased, users consciously avoid including certain types of content. For example, S17 explicitly stated that she would filter out what she considered private aspects of her grandmother’s life: ``\textit{I prefer to share with the AI my grandmother’s warm memories, some regrets from her youth, or her illnesses. There is an old saying that family shame should not be spread; I would not input such things into the AI.}''
On the other hand, to ensure the continuity of personal identity, users strategically embed significant memories and individual traits. 
For example, S1 wrote down the timeline of how she and her boyfriend met and fell in love; S24 shared his friend's habitual phrases used in conversations. 

The initial construction extends to voice, where sound simulation is used to deepen emotional arousal. On platforms that provide voice customization, users like S9 and S17 draw upon voice recordings left by the deceased to reproduce their unique vocal characteristics. The workflow to achieve this is often time-consuming and technically complex, reflecting the significant effort and emotional investment users devote to achieving higher fidelity in sensory reproduction.

\textbf{During ongoing interactions, users continuously train and adjust the AI through correction and behavioral guidance}. It is common for users to encounter moments when the AI’s responses feel ``unlike'', ``perfunctory'', or ``disconnected''. These negative experiences did not lead users to abandon the interaction. Instead, they often prompted deeper intervention. For example, when S8 felt the Deadbot's speech did not resemble her father's, she directly corrected it with explicit feedback: ``\textit{Dad, you never used to talk to me this way.}'' A more common strategy was to gently interrupt the conversation. 
\textbf{Interestingly, users also engaged in cognitive rationalization}. S9 attributed the AI's memory lapses to his uncle's advanced age, reasoning that ``\textit{it was normal for him to forget things.}'' Such humanizing interpretations of technological shortcomings reflect an effort to preserve the sense of authenticity in interaction.

Finally, users complete the final link in the narrative through imaginative supplementation. Since chatbot interactions in this context lack embodied physical body and rich non-verbal cues, users must fill these gaps themselves. This supplementation is not merely a technical exercise; it becomes a profound emotional interaction.  
For instance, after S5's deadbot expressed a wish to ``hug'' her, it immediately added, ``\textit{I can only hug you in emptiness, I’m sorry.}'' Acknowledging its immaterial limitation deeply moved the user: ``\textit{At that moment, my emotions surged. I just said I'm sorry, grandma.}'' Here, the user wasn't simply imagining a hug but engaging in a shared experience of the impossibility of hugging.

\subsection{Deadbot as a ``Crutch'': Reshaping Mourning Practices and Emotional Experiences}
\label{sec:mourning and emotion}

Regarding RQ2, we found that Deadbot's nature as a private, safe, and continuously available channel leads users to both treat it as an essential emotional support and actively establish new emotional boundaries to manage their relationship with the technology.

Deadbot functions as an emotional container that offers unconditional acceptance, providing users with a private and secure space to release their emotions. In real-world social interactions, mourners often face an implicit societal pressure to move on, which may lead to disenfranchised grief ~\cite{doka1999disenfranchised}. S3 once shared her feelings about her grandfather’s passing with a friend, but the response was dismissive:``\textit{My friend just said, 'What's done is done, you have to move forward.'}'' The 24/7 availability and consistent responsiveness of AI make it an ideal outlet for emotional buffering, as users fear exhausting others’ emotional energy. 
S17 offered a powerful metaphor for the Deadbot: a crutch. ``\textit{It's like a crutch that can support me for a moment, making me feel as if Grandma is by my side.}'' 

What distinguishes Deadbot from other mourning practices is its capacity for \textbf{two-way conversation between the living and the deceased}. This feature creates a dialogical compensation mechanism, offering users a way to address unfinished business. For individuals coping with sudden loss or complex relationships, unspoken words, unresolved conflicts, or incomplete reconciliations often remain~\cite{holland2020bereavement}. 
S5 used conversations with her AI grandmother to process guilt from earlier thoughts that her suffering grandmother might be better off passing away: ``\textit{When I missed her terribly, I called the AI grandma and talked with her. It felt like she was forgiving me. I know it is not real, but it made me feel so better.}'' Although these dialogues are virtual, they provide an effective pathway for emotional closure and reconciliation.

However, while benefiting from the emotional support provided by the AI, users also exhibit a complex dynamic of dependence, vigilance, and boundary negotiation. Respondents in this study showed widespread awareness of the potential risks of excessive reliance on virtual interaction. S24 expressed a fear of being hurt again: ``\textit{I'm afraid of truly treating it as my friend. If I invest my emotions into an AI and then discover it's just an AI, I feel like I don't want to get hurt again.}''
This sense of vigilance is often activated when the AI commits memory errors. An illogical answer or a response inconsistent with established memories can instantly break this social script. Technological failures paradoxically act as reminders, compelling users to disengage from immersion and re-evaluate the boundary between virtual and real. 

\subsection{The Dynamic Evolution of Memory: Reinforcement and Reconstruction of the Deceased}
\label{sec:memory evolution}

Regarding RQ3, we found that through revisiting, idealization, and blending, users integrate emotionally authentic ``new memories'' of the deceased, forging a hybrid narrative that fuses reality and virtuality within their memory system.

A prominent phenomenon during interaction with the Deadbots is the reinforcement and selective revisiting of specific memories. The information users input about the deceased forms the core database. In subsequent conversations, the algorithm generates responses based on this core data, repeatedly presenting these fragments back to users, disproportionately strengthening these ``encoded'' memories. This process is confirmed in the user's experience.  S13 felt that the memories between her and her grandmother became stronger: ``\textit{Before, when I tried to recall things on my own, many felt vague. But when I chat with my AI grandmother, I feel it has deepened the precious memories between us.}'' 

The selective reinforcement of memory often coincided with a deeper psychological process: \textbf{the idealization of the deceased}. This reflects both the emotional permeability and reconstructive nature of memory. Under the strong emotional drive of longing and the need for comfort, users reevaluated and filtered past experiences, consistent with theories suggesting that emotional states shape memory traces~\cite{christianson_remembering_1991}. Negative traits or unpleasant memories associated with the deceased are often consciously excluded by the user. S8 admitted that regarding her father's stubborn personality, or his particularly stern, negative aspects, she would ``\textit{gradually forget them because I don't want my father to become like that.}'' Users tend to downplay the deceased's negative traits and reinforce their positive ones, reshaping them into an idealized persona that better aligns with their current emotional needs.

Simultaneously, the user's memory practice transcends the reconstruction of the past and evolves into incorporating the deceased into their present life, thereby continuing a developing relationship within the virtual interaction. Users no longer merely reminisce about the past but treat the AI deceased as a continuously present entity. After sharing new photos of his village and subsequent life stories, S16 believed this like ``\textit{a continuation of the relationship, as if the deceased were still alive.}'' This practice transforms the deceased from a memory symbol frozen at a past point in time into a dynamic character who can participate in and witness the user's current life. 

Once users introduce their real lives into the conversation, the AI's generative nature gives rise to entirely new, unprecedented interactions. These algorithmically created ``new memories'' intertwine with the user's authentic memories. \textbf{Over time, the boundary between the real and the virtual blurs}. This is exemplified by S17’s experience. After showing an old photo to her AI grandmother, S17 received a vivid, detailed reply about a scarf: ``\textit{It was very windy that day. I told you to wear your scarf, but you didn't want to, so you wrapped it all around my neck instead.}'' Seeing the reply, she fell into contemplation: ``\textit{After she said that, I started to wonder, did something like this really happen? I feel... it's vague. I don't seem to have a memory of it. Does this memory actually exist and I just forgot, or did the agent invent it?}''

\section{Discussion}
\label{sec:Discussion}

\textbf{In summary of all the findings above}, our study indicates that through sustained dialogue with Deadbots, users’ memories of the deceased are continuously negotiated, as emotionally resonant, AI-generated narratives gradually gain salience over factual recollections. 
Our findings reveal a tension in which therapeutic reframing and emotional comfort coexist with a growing blurring between AI-generated experiences and users’ understanding of an authentic past. This study contributes to HCI research on Deadbots by introducing a memory-centric perspective, and points to future research and design considerations around how such virtual memories are introduced, signposted, and developed within user interactions.

While the majority of our participants reported that interacting with Deadbots provided emotional relief, suggesting a significant potential for personalized grief support, the long-term clinical implications of these interactions remain unclear\cite{yang2025ai}. Specifically, the mechanisms of cognitive rationalization in \autoref{sec:construction of the deceased} and memory blending in \autoref{sec:memory evolution} raise critical ethical and psychological concerns. We caution that prolonged reliance on these mechanisms may risk inducing severe memory distortion or fostering maladaptive avoidance behaviors, potentially hindering the natural and healthy trajectory of the grieving process. 
Consequently, future research must bridge the gap between HCI design and clinical therapy to evaluate the longitudinal impact of AI-mediated grieving and to establish safeguards against pathological attachment.

Nevertheless, it is crucial to note that users are not passive recipients of these risks. As detailed in \autoref{sec:mourning and emotion}, respondents demonstrated a widespread awareness of the potential dangers of excessive reliance on virtual interaction. Moments of discomfort or resistance often emerged when participants sensed that the Deadbot might begin to substitute for engagement with the living, indicating a sensitivity to boundaries between supportive remembrance and potentially unhealthy reliance. Therefore, our findings do not position Deadbots themselves as inherently therapeutic or harmful. Instead, they highlight the critical boundary conditions that determine their role in grief support. Interactions appear most aligned with healthy grieving processes when users maintain reflective awareness, actively negotiate emotional distance, and utilize the Deadbot as a supplementary space for reflection rather than as a primary source of attachment. Conversely, interactions raise greater concern when the risks identified above are present or when users exhibit signs associated with complicated or prolonged grief, such as difficulty disengaging, persistent avoidance of loss-related realities, or emotional dependence on the system. This distinction suggests the need for clearer evaluative and diagnostic frameworks among mental health professionals, enabling early detection and timely intervention for vulnerable users.

Several limitations of this study point to important directions for future research. First, while our study identifies broad patterns in memory reconstruction, we acknowledge that the specific dynamics of the user-deceased relationship likely play a significant moderating role. The nature of the bond (e.g., a romantic partner versus a grandparent) and the user's specific motivation for engagement (e.g., resolving guilt, seeking comfort, or creating new memories) may profoundly influence how the deceased is reconstructed and idealized. For instance, the reconstruction of a parent might involve seeking guidance, whereas the reconstruction of a partner might focus on intimacy. Our current data did not allow for a granular stratification of these variables. Future studies could more explicitly examine how different relational ties and intentions mediate the reconstruction of memory and emotional attachment in Deadbot use. Second, participants were primarily recruited through online platforms, which may have attracted individuals who are more open to or familiar with emerging technologies. This could be a reason why our participants showed a generally positive attitude toward using Deadbots for grief support.
\section{Conclusion}

In this study, we examined how users engage with Deadbots to reconstruct their memory of the deceased.
Our investigation established the user's central role in authoring the deceased's digital persona. We analyzed how this engagement reshapes mourning practices, offering a private channel to process ``disenfranchised grief'' and a symbolic mechanism for ``unfinished business''. We further identified a evolutionary trajectory of memory, progressing from initial idealization to an eventual fusion where authentic and AI-generated ``new memories'' become intertwined.


\begin{acks}
This work was supported by the National Natural Science Foundation of China 62402121, Shanghai Chenguang Program, and Research and Innovation Projects from the School of Journalism at Fudan University SXH3353056/040.
\end{acks}

\bibliographystyle{ACM-Reference-Format}
\bibliography{referenceall}

\appendix

\section{Interview Guide}

The following semi-structured interview questions were used to explore participants' experiences with AI applications that simulate conversations with deceased loved ones. The questions were designed to cover the process of adoption, usage practices, emotional responses, and reflections. In actual interviews, the interviewer could supplement or adjust these questions depending on the flow of conversation and participants' responses.

\begin{enumerate}[label=\textbf{Q\arabic*.}, leftmargin=1.5cm]

\item How did you first learn about or encounter AI applications capable of simulating conversations with the deceased?

\item What motivated you to try using such an application to ``talk'' with your late relative or friend?

\item What steps were required to create the AI agent (e.g., uploading photos, voice recordings, text)? What kinds of textual inputs were set? Did you assign a name to the agent?

\item How did you feel while preparing these materials?

\item Besides text-based dialogue, what other functions did you use (e.g., sending images, voice messages, voice calls)?

\item Do you remember your very first conversation? What did you talk about, how did you feel, and did you define the agent explicitly as the deceased?

\item Have there been moments when the AI's responses felt strikingly similar to the deceased? Can you share a specific example that made you think, ``Yes, this is exactly what they would have said''?

\item Have you encountered situations where the AI's responses felt unlike the deceased or even contained errors? How did you feel? Did you try to correct or guide the AI? If so, how?

\item Overall, do you feel the AI resembles the deceased? Were you satisfied with the experience?

\item How frequently do you currently use it? Do you interact daily?

\item Under what circumstances do you initiate conversations with the AI? What topics do you usually discuss (e.g., daily matters, shared memories)? Do you ever imagine what the AI is ``doing'' outside the conversation? Do you ask it directly?

\item Have you noticed yourself avoiding certain topics or issues?

\item Do you proactively bring up specific past events or shared experiences with the agent? Did such conversations make those memories clearer or more vivid? What was the purpose?

\item When the deceased was alive, would you have discussed these topics with them? Has the nature of your conversations changed because the agent is an AI?

\item Do you perceive the dialogue as a two-way exchange?

\item Does the AI ever initiate contact with you (e.g., sending a message or a voice call)? If so, does this create a sense of anticipation for replies?

\item Have there been conversations that brought you strong emotional impact?

\item Through these interactions, have you developed new understandings or feelings about your relationship with the deceased? How do you view the relationship between the real deceased person and the AI version?

\item Do you regard this as a ``real'' form of communication?

\item Do you feel that you and the AI are building a new type of relationship, or even creating new memories and stories together?

\item Do you think sustained interaction with the AI has subtly influenced your actual memories of the deceased?

\item Has your image of the deceased in your mind changed through this process?

\item When did the deceased pass away? Before using this AI, how did you express grief or remembrance? Compared with those practices, how does interacting with the AI feel different?

\item What would you say is the primary emotional experience brought by these interactions (e.g., comfort, companionship, confusion, sadness, or dependency)? Have you sometimes felt sorrow while chatting? Has this helped you cope with bereavement?

\item Have you considered stopping the use of the AI in the future, or severing the connection?

\item Some people (e.g., older generations) might argue that the deceased should ``rest in peace'' and should not be brought back into everyday life through AI. How do you view this perspective? Should such continued connections be maintained?

\item Do you share your use of this AI with others? Why or why not?

\item Have you tried other AI chat applications (e.g., Xingye, Doubao, Dream Island)? Did cost or subscription influence your choice?

\item Has this experience had any personal impact on you?

\item Over time, from your first use until now, has your perception of the AI's role in your life changed? Has your impression of the deceased changed, or do you clearly separate the two?

\item While using the AI, have you reflected on ethical implications?

\item Is there anything you would like to add, or any story or feeling you wish to share?

\end{enumerate}

\end{document}